\documentclass[%
 reprint,
 amsmath,amssymb,
 aps,
]{revtex4-2}

\usepackage{graphicx}
\usepackage{dcolumn}
\usepackage{bm}
\usepackage{color}
\setcitestyle{super}

\begin{document}

\preprint{APS/123-QED}

\title{Charge dynamics in spintronic THz emitters}
\author{Georg Schmidt}
 \email{georg.schmidt@physik.uni-halle.de}
\affiliation{Institut f\"ur Physik, Martin-Luther-Universit\"at Halle-Wittenberg, Von-Danckelmann-Platz 3, 06120 Halle, Germany \\ Interdisziplin\"ares Zentrum für Materialwissenschaften,
Martin-Luther-Universit\"at Halle-Wittenberg, Heinrich-Damerow-Strasse 4, 06120 Halle, Germany
}
\author{Bikash Das-Mohapatra}
 \affiliation{Institut f\"ur Physik, Martin-Luther-Universit\"at Halle-Wittenberg, Von-Danckelmann-Platz 3, 06120 Halle, Germany}
\author{Evangelos Th. Papaioannou}
 \affiliation{Institut f\"ur Physik, Martin-Luther-Universit\"at Halle-Wittenberg, Von-Danckelmann-Platz 3, 06120 Halle, Germany}

\date{\today}

\begin{abstract}
We show that to correctly describe the ultrafast currents in spintronic THz emitters it is necessary to take charge equilibration into account. The charge current which is locally induced by a fs laser pulse and the inverse spin-Hall effect (ISHE) leads to ultrafast charging phenomena at the edge of the illuminated area. Subsequent discharging leads to a current backflow with a delay and a time constant that mainly depends on the conductivity of the emitter. On the one hand, only this delayed charge equilibration allows the detection of the primary current pulse via THz emission because an instantaneous backflow would cancel any far field emission. On the other hand, especially for longer light pulses the backflow can significantly change the emitted spectrum compared to the initial spin-current pulse by suppressing low frequency components. For the analysis of spin physics based on the charge current profile it is important to understand that the timing of the spin current cannot be inferred from the charge current unless the contribution by the ISHE and the backflow can be deconvoluted.

\end{abstract}

\maketitle


Spintronic THz emitters (STE) is a new field of research that has been heavily investigated over the past years\cite{kampfrath_2013,seifert,Huisman,Munze2016,ADMA:201603031,Torosyan2018,Papa2018,nenno_2019,Jaffres2020,MPaBe,ThomasThomson2021,Oliver2021,TanWei2021,Khusyainov2021,Gupta2021,seifert3}. In typical experiments a fs laser pulse hits a bilayer consisting of a ferromagnet and a heavy metal. The laser pulse induces a spin current from the ferromagnet into the heavy metal which therein is converted into a lateral charge current by the inverse spin Hall effect (ISHE). This ultrafast current pulse leads to the emission of THz radiation.
From the detected THz signal the timing of the current pulse can be extracted. Typically, the temporal shape of the current pulse is then used to analyze the ultrafast spin physics in the STE under illumination.

The analysis, however, is based on the assumption that the charge current is directly proportional to the spin current \cite{ADMA:201603031,nenno_2019,kampfrath_2013,seifert} $j_{charge}(t)=\Theta j_{spin}(t)$ with $j_{charge}$ the charge current, $\Theta$ the spin Hall angle, and $j_{spin}$ the spin current.
In these evaluations the current pulse often consists of a positive peak followed by a negative one, a fact that has for example been explained by a leading contribution caused by the majority spins entering the heavy metal followed by a trailing negative component caused by the minority spins\cite{kampfrath_2013,nenno_2019}. Up to now, little attention has been given to the dynamics that are simply the result of charge diffusion, electric fields and the geometry of the emitter and the laser spot.

The primary charge current that is a direct result of the conversion of the spin current displaces charge. When the spin current is over, diffusion and local electric fields within the Debye screening length must restore the condition of charge neutrality $\rho(\vec{x})=0$ before the local charge current goes back to zero. As a consequence, the shape of the current pulse must not be explained without taking into account the current dynamics in the metal film caused by the restoration of charge neutrality. This is even more important as the far field detection of THz radiation does not give access to the exact current distribution in the emitter. In the following, we investigate theoretically a gedankenexperiment in which the laser pulse induces a positive current pulse with a simple Gaussian time dependence for different laser spot sizes, pulse lengths, thicknesses, and STE conductivities, respectively. We then model the subsequent charge redistribution and show the resulting time dependence of the equally ultrafast current backflow.

The structure under investigation consists of a square metal sheet of $l\times w=150\times150\,\mu m^2$. The length and width are chosen to allow the approximation of an infinite size STE compared to the laser spot.
Emitter and laser spot (radius r) are centered at $x=y=0$ and a constant laser intensity over the circle is assumed. Although this neglects a typical beam profile this distribution is sufficient to demonstrate the underlying physics. For the time dependence of the intensity we chose a gaussian peak with a full width at half maximum of $\tau$.

For the simulations the Comsol electric current module was used and the material parameters required were  Electrical Conductivity($\sigma$) and Relative permittivity ($\epsilon_{r}$). For Sapphire substrate layer $\sigma=10^{-12}$ S/m and $\epsilon_{r}=3.064${\cite{harman}}. For the STE thin film we have taken $\sigma=5\times10^{6}$\, S/m{\cite{seifert,seifert2,keller}}, which is in good agreement with the conductivity values for multilayer thin films with W, CoFeB and Pt and for simplification an average $\epsilon_{r}$ of 903 (Pt)\cite{ordal}. Our reference STE (dubbed $STE_{ref}$ in the following) has $r=10\,\mu m$,  $\tau=20\,fs$, and an STE thickness of t=10\,nm.

\begin{figure}
    \includegraphics[width=0.9\linewidth]{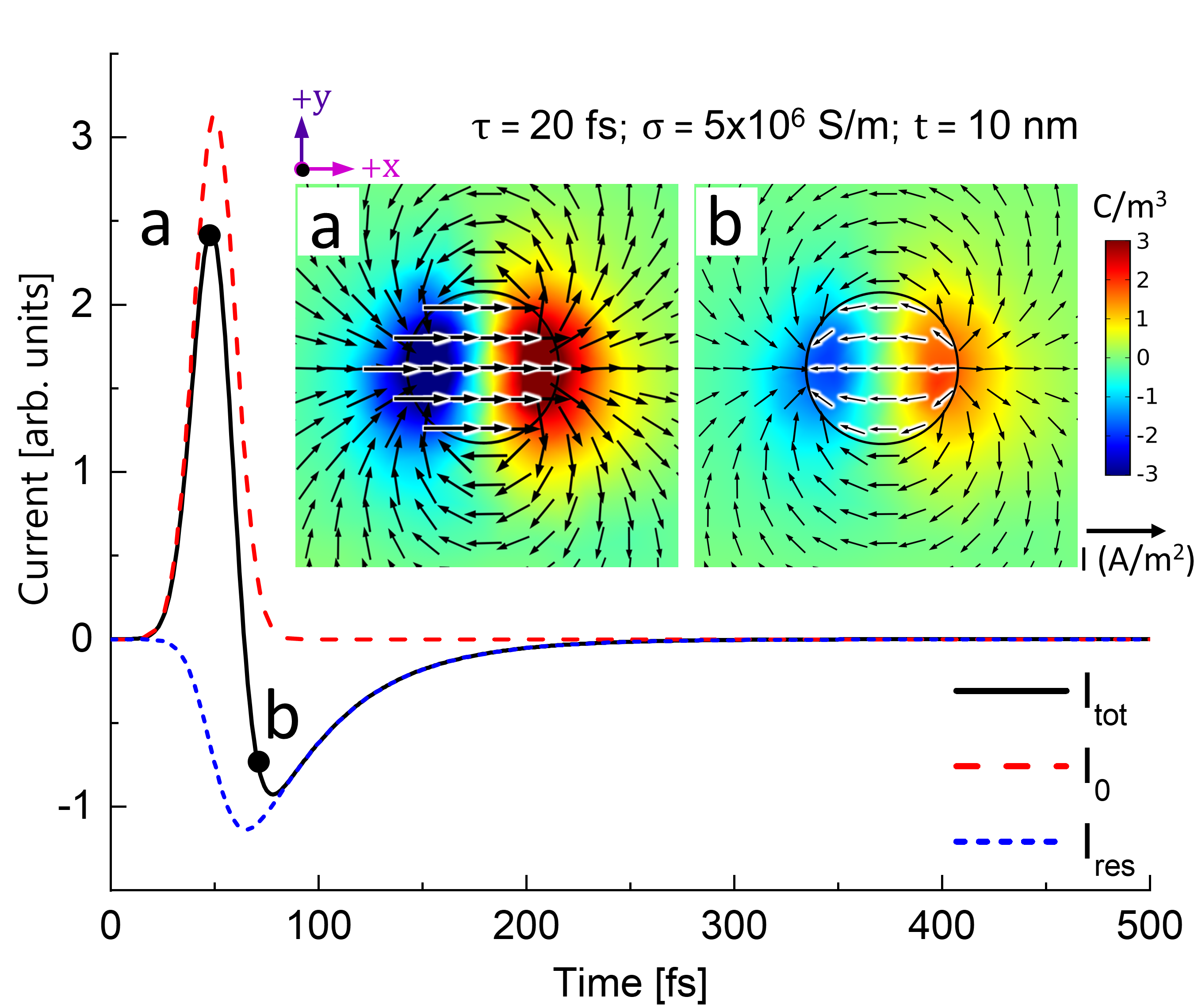}
    \caption{Current over time for the original excitation pulse $I_0$ (dashed red), the total current $I_{tot}$ (solid black) and the response of the system $I_{res}$ (dotted blue) for a 10 nm thick emitter with a conductivity of $5\times 10^6 S/m$. The excitation pulse has a FWHM of 20\,fs and is centered around t=100\,fs. Insets show the current and charge distribution at two different times, respectively, marked a and b in the black curve.\label{fig:1}}
\end{figure}

Via spin current and the ISHE the light creates a lateral current density $j_0$ whose direction we define as +x. For sake of simplicity $j_0$ is set proportional to the local light intensity. We implement this by adding a uniform $j_0(t)$ inside the circular spot. To leave the system free to react to the imprinted current in every place, we don't set any boundary conditions for the simulation in terms of potentials, fields, or current densities. To compare our results with THz emission experiments it is crucial to understand that the far field emission for a small STE reflects the integrated current but not the local current density. We thus integrate the local current density over the whole emitter for each moment in time that we are investigating:

\begin{equation}
    \vec I(t)=\int_{Volume}\vec j(\vec x,t) d\vec x
\end{equation}

The total current can be written as $\vec I_{tot}(t)=\vec I_0(t)+\vec I_{res}(t)$ with $\vec I_0(t)$ being the integrated ISH-current and $\vec I_{res}(t)$ the system response. It is self evident that even a very large emitter is still a closed system in terms of charge conservation and the total current  integrated over time must be zero. For the two respective parts $\vec I_0(t)$ and $\vec I_{res}(t)$, however, this is not the case because the ISHE current leaves part of the charge displaced and the system locally charged, while the system response restores charge neutrality from a charged state. Locally, the charging caused by a current density is determined by the continuity equation $\vec\nabla\vec j=-\frac{\partial\rho}{\partial t}$.

Because we allow for transient charging $\vec\nabla\vec j$ can become finite on very short timescales. Integrated over time, however, the assumption of charge neutrality in the steady state requires that locally the condition

\begin{equation}
    \int_{-\infty}^{\infty}\vec\nabla\vec j(x,t) dt=0
\end{equation}

\noindent is fulfilled.

Fig. \ref{fig:1} shows the current/time dependence for $STE_{ref}$. The red dashed curve shows the integrated current density $\vec I_0(t)$ induced by the laser and by the ISHE that we impose on the system. We call the corresponding current density $\vec j_0(\vec x,t)$. The black solid curve $\vec I_{tot}(t)$ shows the total integrated current density that is the sum of the current caused by the ISHE and the system response (this density is called $\vec j_{tot}(\vec x,t)$). The blue dotted curve only shows the system response $\vec I_{res}(t)$ or the integral over the response current density $\vec j_{res}(\vec x,t)$.

\begin{figure}
    \includegraphics[width=1.02\linewidth]{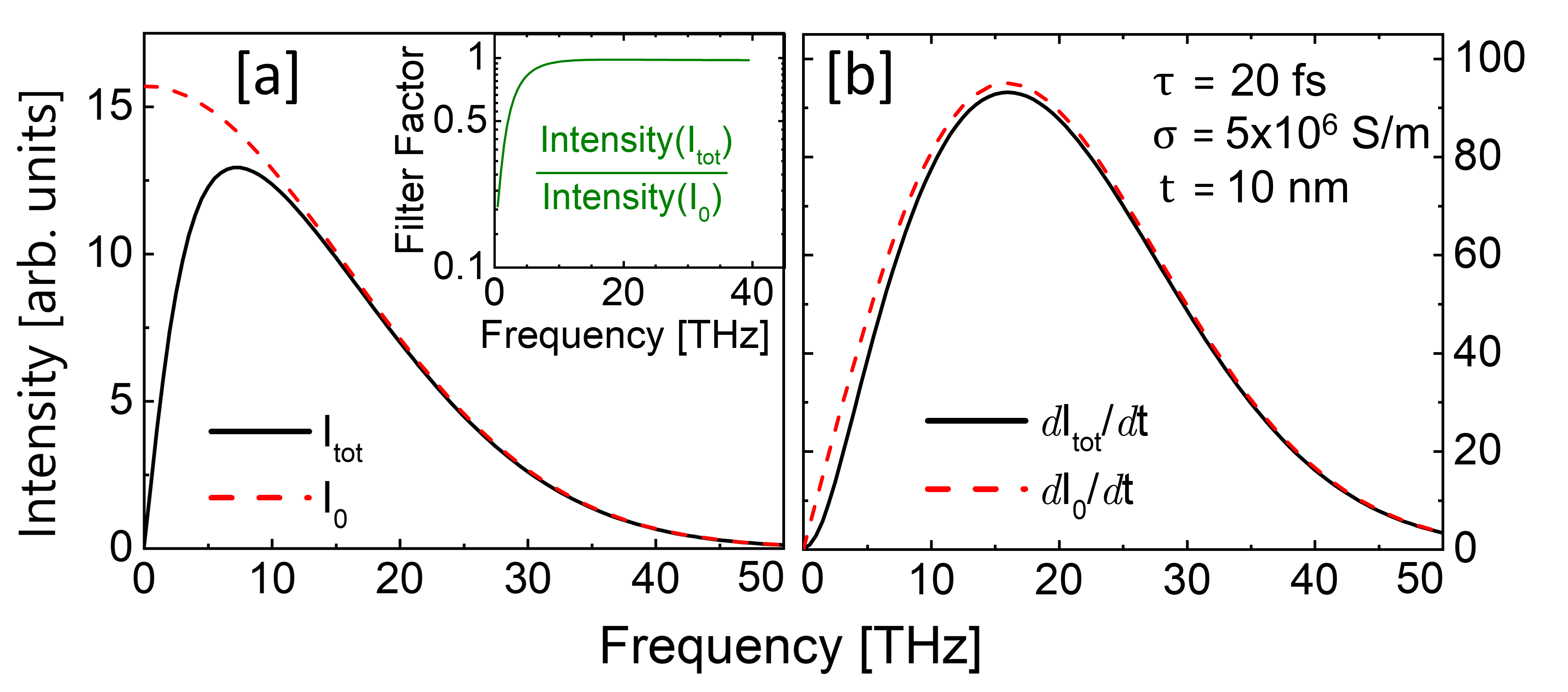}
    \caption{ Spectral intensity for current (a) and for its time derivative (b). At high frequency the spectral intensity is identical for both signals while lower frequencies that are present in the excitation $I_0$ are suppressed in the total signal $I_{tot}$. The inset of (a) shows the ratio of the spectral intensity for $I_{tot}$ and $I_{0}$ plotted over frequency\label{fig:2}}
\end{figure}

In the beginning of the pulse we see $\vec I_{tot}(\vec t)$ rising with the slope of $\vec I_{0}(t)$. After a very short time the system response starts and a negative contribution by $\vec I_{res}(t)$ appears. As we can see this contribution is asymmetric around its peak and the peak is delayed with respect to the peak of $\vec I_{0}(t)$. As a consequence, several things can be observed. Firstly the maximum of $\vec I_{tot}(t)$ is smaller than that of $\vec I_{0}(t)$. Secondly the decay of $\vec I_{tot}(t)$ after the maximum is faster than would be expected from the shape of $\vec I_{0}(t)$. Finally, the positive part of $\vec I_{tot}(t)$ is followed by a negative part, when $\vec I_{0}(t)$ is almost at zero and the backflow  restores the charge neutrality.

These curves can be understood taking into account local charging $\rho(x,t)$ and current densities $\vec j_{tot}(\vec x,t)$ at different times of the experiment (inset of Fig. \ref{fig:1}). The current density is expressed by arrows where the direction of an arrow shows the direction of $\vec j$ while its length is proportional to $|\vec j|$.

Insert a shows current and charge distribution at the maximum of $I_{tot}$ which is slightly earlier than the maximum of $I_0$. We can see a massive charging on the left and right hand side. The current inside the spot is mainly determined by $I_0$ while on the outside already a considerable backflow ($I_{res}$) is visible. It should be noted that also inside the spot there is a contribution by $I_{res}$ which is, however, masked by the large value of $I_0$. At point b, $I_0$ is back to zero. Nevertheless, still a charging remains and we now observe a backflow $I_{res}$ inside and outside the spot, resulting in $I_{tot}<0$.

This current distribution can be understood as follows: We start with only current in +x direction within the spot. Inside the spot, charge is displaced but no charge accumulates. Outside the spot we have $j=0$. Only at the circumference we have $\vec\nabla\vec j=\partial{\rho}/\partial{t}\ne 0$ resulting in local charging.
With current along the x-direction, charging is maximised at y=0 and x=$\pm$r and minimized for x=0 and y=$\pm$r. So positive charge starts to accumulate on the +x side while negative charge is left on the -x side. This charging can be intuitively understood when taking into account that it is a basic principle of electrodynamics that every part of the metal sheet has some capacitance that, although very small, can be charged and is discharged via any connecting material with finite conductivity. The small capacitance of metal microstructures is for example well known and extensively used in coulomb blockade experiments. The related time constants, however, are barely observed in sub THz electronics.
$I_0$ charges the capacitance at the edge of the spot. When the charge builds up, it also immediately starts to flow out; the capacity is discharged. Discharge happens more or less symmetrically around the respective charge accumulations. As a result we have a backflow around and within the spot. At first, the latter, however, is not large enough to completely reverse the current within the spot. While the backflow around the spot persists almost from the beginning of the excitation pulse, a complete reversal of the current inside the spot only happens when $\vec I_{tot}(t)$ is also getting negative. It should be noted, however, that at this point $\vec I_{0}(t)$ is still positive, so the backflow overcompensates the initial excitation.
One detail should be pointed out: The currents within the spot but also the backflow outside the spot are at a maximum at the maximum of $\vec I_{tot}(t)$, an effect that might be expected. The charge density, however, reaches its maximum a bit later when $\vec I_{tot}(t)$ is already decreasing. More figures of current and charge distribution at different times can be found in the supplemental material.

The trailing negative current peak also modifies the emitted THz spectrum. Fig.\ref{fig:2}a shows the FFT $\hat{I}$ for the current $\vec I_0(t)$ and for $\vec I_{tot}(t)$. As we can see there is a strong suppression of lower frequencies. With increasing frequency the spectral amplitude for $\vec I_{tot}(t)$ approaches that of $\vec I_{0}(t)$ until they completely coincide for frequencies higher than approx. 15 THz. This behavior is explained by the equivalent circuit in Fig.\ref{fig:4} that shows that the structure acts as a high pass filter.
Because the THz emission is proportional to the time derivative of the current and not the current itself we also show the fourier transform of $dI/dt$ (Fig.\ref{fig:2}b). The fourier transform of the time derivative $dI/dt$ is proportional to $\omega \hat{I}$ and thus already has a reduced intensity at low frequencies making the effect appear less pronounced. To visualize the filter function, we divide the spectral intensity of $I_{tot}$ by that of $I_0$. The result are high pass characteristics with a suppression of a factor of 2 or more below 3 THz. The original level is only obtained for f$>$10 THz.
As an important test we vary the spot size that might play a role if the two charge accumulations interacted and behaved as a dipole. Like in a Hertzian dipole a larger spot would lead to a lower frequency or at least a slower response. The simulations, however, show no influence. For spot sizes of  $r=25\,\mu m$ and $r=2.5\,\mu m$ all cures are simply scaled in magnitude by constant factors with respect to those for $r=10\,\mu m$ while the timing remains.

\begin{figure*}
    \includegraphics[width=18cm]{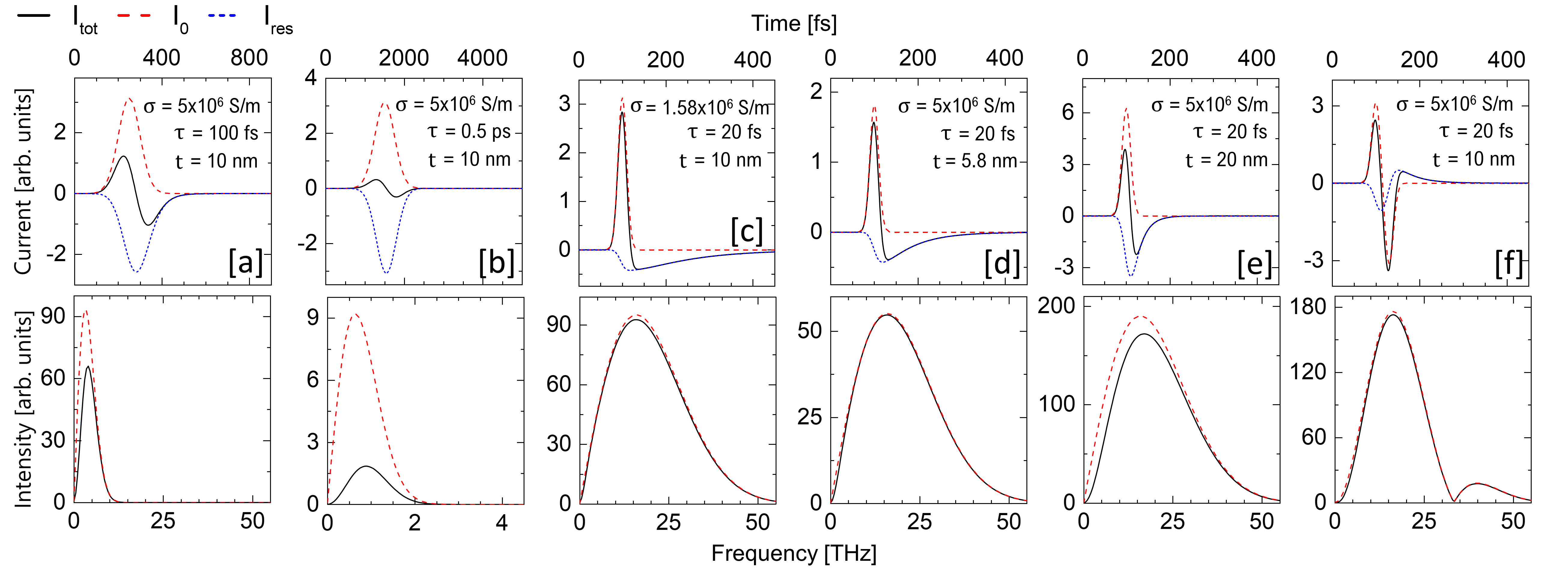}
    \caption{Time domain signals and spectra for six different scenarios. (a) and (b) show the results for a 100 fs wide and a 0.5 ps wide pulse $I_0$, respectively using $STE_{ref}$. (c) Shows the same experiment as in fig.\ref{fig:1} using $STE_{ref}$ with a conductivity reduced to $1.58\times 10^6$. The increased time constant of $I_{res}$ (blue dotted) reduces the cutoff frequency making the impact on the spectrum very small. (d) shows results for $STE_{ref}$ with a reduced thickness (5.8\,nm). The increased area resistance has a similar effect as in (c). Decreasing the resistance by making $STE_{ref}$ thicker (e) shifts the cutoff frequency to higher values and decreases the signal in the range up to more than 25 THz. (f) shows the result for an antisymmetric pulse using $STE_{ref}$. Because this pulse has less low frequency contributions the effect on the lower part of the spectrum is less visible but still a reduction is observed up to almost 20 THz.\label{fig:3}}
\end{figure*}

\begin{figure}
    \includegraphics[width=4cm]{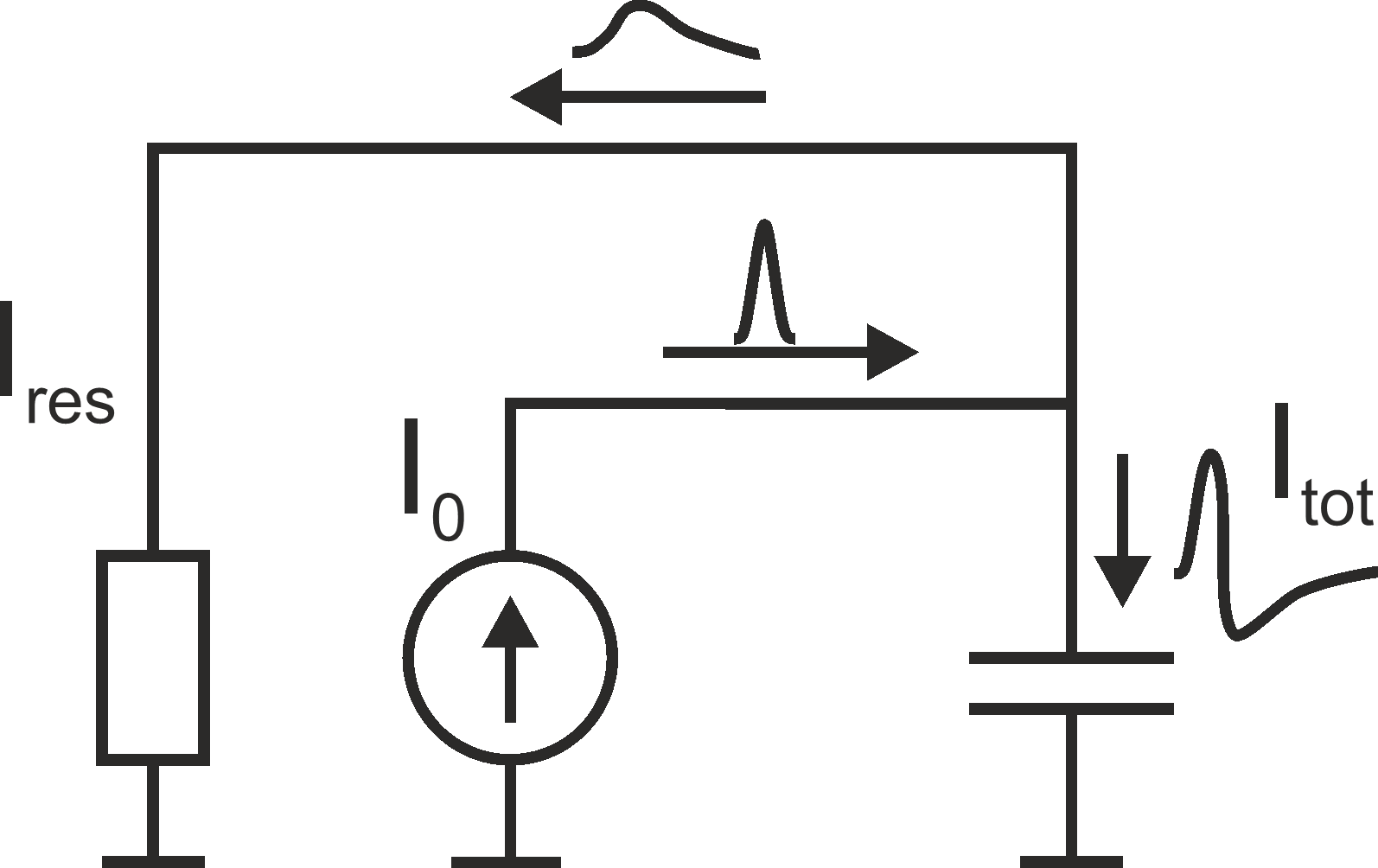}
    \caption{Equivalent circuit for the STE. The current pulse charges the capacitor that can then discharge with a certain time constant through the resistor.\label{fig:4}}
\end{figure}

Changing the duration of the laser pulse from $\tau=20\,fs$ (figs. 1 and 2) to $\tau=100\,fs$ (fig. \ref{fig:3}a) reflects the physics behind the system's response. It must be understood that counter intuitively, only the finite R/C time constant of the system allows the positive current peak to exist. Without the system capacitance $\vec I_{0}(t)$ and $\vec I_{res}(t)$ would be simultaneous and $\vec I_{tot}(t)$ would be 0 with no THz emission. With 100 fs the time constant of the excitation pulse is longer, while the R/C constant of the emitter remains the same. On the timescale of $\vec I_{0}(t)$ the backflow now starts earlier and rises faster. In the spectra, the total intensity is much more reduced from $I_0$ to $I_{tot}$ than for the 20 fs pulse. However, this is merely a result of the more abundant lower frequency components due to the longer pulse. The relative suppression at the respective frequencies is the same as for the 20 fs pulse because the material parameters and thus the filter characteristics were not altered. To emphasize the effect we also show a simulation for a peak with 0.5 ps FWHM (fig. \ref{fig:3}b). As the spectra show, $\vec I_{0}(t)$ would still have considerable contributions up to more than 1 THz. The backflow, however, more or less completely annihilates this frequency regime.

Now, we change the R/C time constant of the emitter while keeping the pulse length at 20 fs. First we reduce the conductivity to $1.58\times10^6\,S/m$ (fig. \ref{fig:3}c). From the decay of $\vec I_{res}(t)$ that is determined only by the R/C time constant we can see that the discharge is now slower. At the same time the lower frequencies are less suppressed than for the original conductivity corresponding to a lower cutoff frequency of the filter characteristics (The filter characteristics of all emitters are shown in the supplementary material).

A change in layer thickness changes the in-plane conductivity while keeping the capacitance virtually unchanged. For a 5.8 nm film we observe a larger time constant (fig. \ref{fig:3}d) than for the original emitter while for 20 nm (fig. \ref{fig:3}e) the time constant decreases, raising the cutoff frequency. Here at 5 THz the amplitude is still suppressed by almost a factor of 2 and even at f=20 THz the intensity is only back to 92\% of the original value. For all different emitters the filter characteristics are shown in the supplementary material.

Because a simple gaussian profile oversimplifies the ultrafast spin current we also give an example for a more complex excitation (fig. \ref{fig:3}f). With an STE of 20 nm thickness  we use for $I_0$ a sequence of two identical gaussian pulses with opposite sign that form a bipolar antisymmetric shape. The resulting $\vec I_{tot}(t)$ looses this symmetry and even has a small positive trailing pulse. The modification to the spectrum is similar to (fig. \ref{fig:3}e) because the emitter is the same. Please note that a current pattern inferred from existing experiments with only one positive and one trailing negative component must be caused by a single positive spin current pulse while a bipolar excitation leads to a more complex signal.

We can describe our system by a relatively simple equivalent circuit (fig.\ref{fig:4}). The ISH-current $\vec I_{0}(t)$ can be considered as a current source that charges a capacity. The capacity is discharged via the resistance of the surrounding material leading to the current $\vec I_{res}(t)$. Only $\vec I_{tot}(t)$, the sum of the two is relevant for the THz emission because in the far field, radiation created by the different components though not coming from exactly the same place cannot be distinguished whereas a near field detector might be able to discern these signals.
The current in this system is described by the simple differential equation:

\begin{equation}
    C\frac{dU(t)}{dt}=I_0(t)-\frac{U(t)}{R}
\end{equation}

This equation describes a low order high pass filter. High frequencies pass the capacity unimpeded while lower frequency currents are more and more flowing via the resistance R. As a rule of thumb: A metal disc of $r=10\,\mu m$ has a capacitance of approx. $1\,fF$ so discharging through a resistance of $R=10\Omega$ leads to a time constant of $\tau=10 fs$.

This model also has a simple consequence for deriving the charge current pattern from a  THz emission experiment. Any result must obey

\begin{equation}
    \int_{-\infty}^{\infty}\vec I_{tot}(t) dt=0
    \label{TotalZero}
\end{equation}.

Any current pattern not doing this violates the laws of electrodynamics allowing for a consistency check of the result.

It is clear that the model used here includes a number of simplifications, for example by neglecting inductive effects. Nevertheless, this does not alter the underlying physics and will only add minor corrections. Nevertheless it has to be kept in mind if one tries for example to extract exact STE parameters from a real experiment.

As a consequence, it is not allowed to extract the spin current from the charge current inferred from the THz signal without taking the charge redistribution into account that significantly changes the profile in time. This effect complicates the extraction of the original spin current from the detected THz radiation and may lead to misinterpretation of the underlying spin physics. A deconvolution of the different mechanisms is further complicated by the fact that the laser spot profile is not uniform but has at least a radial intensity profile. The time constant of the system reduces the lower frequency part of the spectrum but not the higher frequencies and the corresponding cutoff frequency becomes higher with increasing conductivity of the emitter. Especially for a measurement setup with an upper frequency limit between 5 and 10 THz this may give the impression of a reduction of the total THz emission while in fact higher undetected frequencies are not reduced.
As an outlook, it should be noted that the design of the STE can be used to shape the current pulse. Especially and quite counter intuitive, a larger time constant of the system will reduce the suppression of the lower frequency part of the spectrum and broaden the spectral width towards lower frequencies.

We acknowledge the support of the Deutsche Forschungsgemeinschaft in TRR227 TP B02.

\end{document}


\preprint{APS/123-QED}

\title{Supplementary material for: Charge dynamics in spintronic THz emitters}
\author{Georg Schmidt}
 \email{georg.schmidt@physik.uni-halle.de}
\affiliation{Institut für Physik, Martin-Luther-Universität Halle-Wittenberg, Von-Danckelmann-Platz 3, 06120 Halle, Germany \\ Interdisziplinäres Zentrum für Materialwissenschaften,
Martin-Luther-Universität Halle-Wittenberg, Heinrich-Damerow-Strasse 4, 06120 Halle, Germany
}
\author{Bikash Das-Mohapatra}
 \affiliation{Institut für Physik, Martin-Luther-Universität Halle-Wittenberg, Von-Danckelmann-Platz 3, 06120 Halle, Germany}
\author{Evangelos Th. Papaioannou}
 \affiliation{Institut für Physik, Martin-Luther-Universität Halle-Wittenberg, Von-Danckelmann-Platz 3, 06120 Halle, Germany}

\date{\today}

\maketitle


During and well after the initial spin current pulse the current back flow starts. This becomes first visible in a back flow around the excitation spot. Lateron even the current inside the spot is compensated and then fully reversed. Fig. 1 shows the characteristics of current over time in $STE_{ref}$ with a number of times marked and labeled. In the main manuscript current and charge distribution were shown for two characteristic times. In Fig. 2 current and charge distribution are shown for each of the 10 different times, respectively for the interested reader. It is nicely visible that already at point g the total current $\vec I_{tot}$ is at zero because the back flow outside compensates the still positive current inside the spot. At time h, even the current inside the spot is reversed and $\vec I_{tot}$ becomes negative,  although $\vec I_0$ is still finite and positive. Even 100\,fs after the initial pulse, the total current is still negative and has not yet decayed to zero.

\begin{figure*}
    \includegraphics[width=0.9\linewidth]{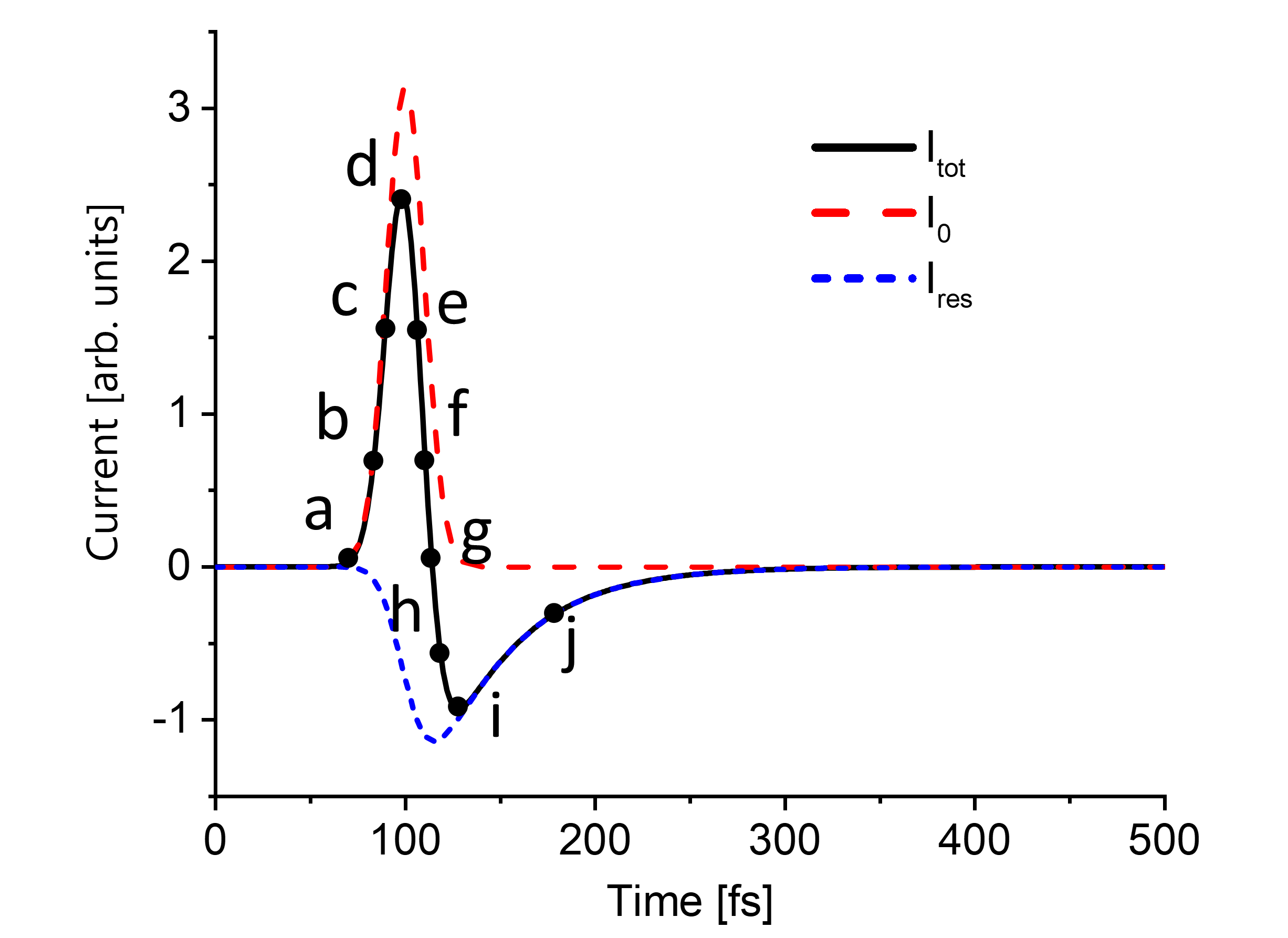}
    \caption{Current over time for  the original excitation pulse $I_0$ (dashed red), the total current $I_{tot}$ (solid black) and the response of the system $I_{res}$ (dotted blue) for the reference experiment with $STE_{ref}$ (10 nm thick emitter with a conductivity of $5\times 10^6 S/m$, $\tau=20\,fs$). The dots labeled a-j indicate 10 different times for which fig. 2 shows the respective charge and current distribution.\label{fig:1}}
\end{figure*}

\begin{figure*}
    \includegraphics[width=18cm]{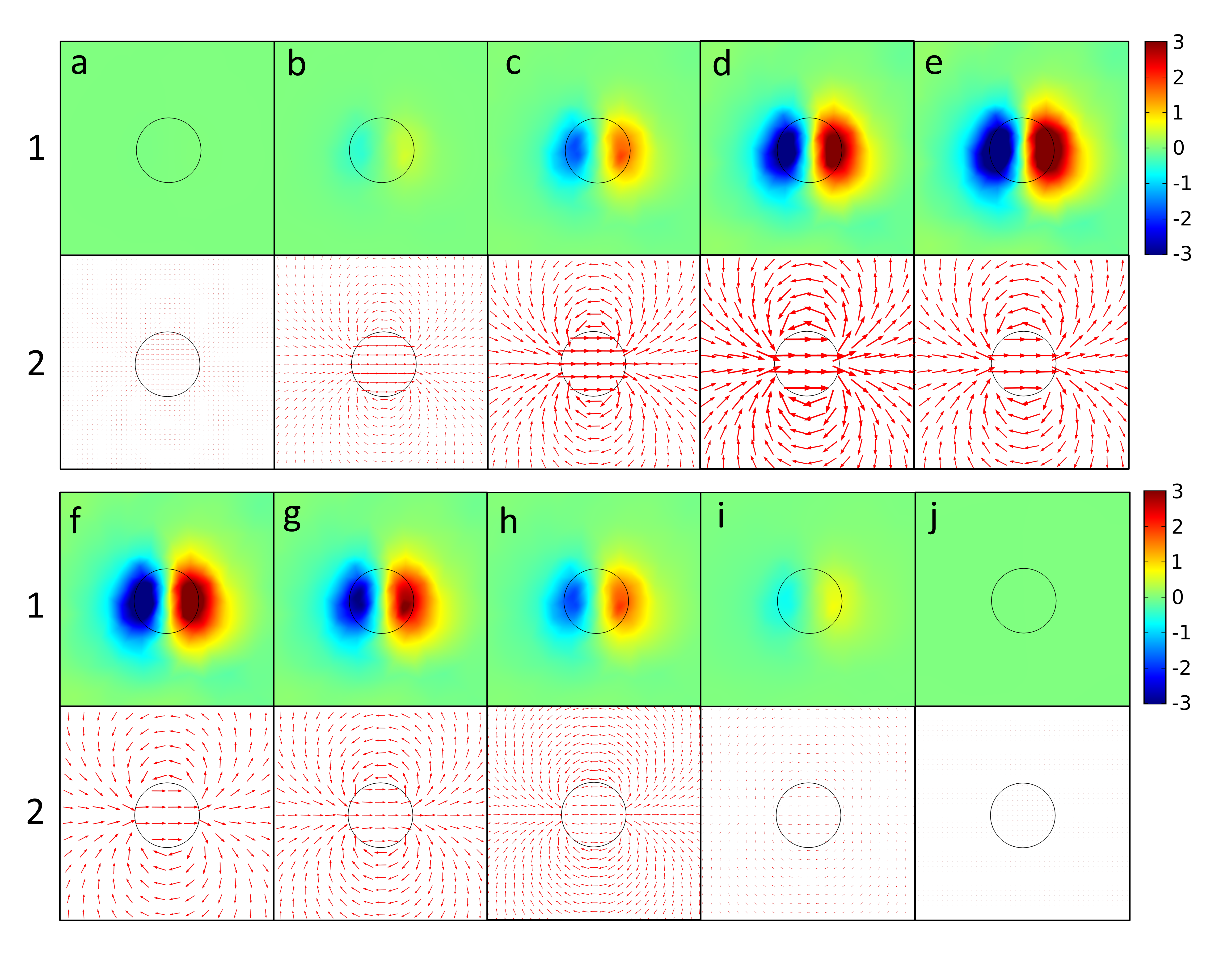}
    \caption{a-j show the charge (row 1) and current (row 2) distributions for the 10 different respective times marked in fig. 1.\label{fig:2}}
\end{figure*}

As shown in Fig. 2 of the main manuscript one can calculate filter characteristics for the different emitters by dividing the spectral intensity of the original pulse's derivative $d\vec I_{tot}/dt$ by that of the total current $d\vec I_0/dt$. This way we obtain high pass filter characteristics that reduce certain parts on the low frequency side of the spectrum, depending on the conductivity of the STE. Fig. 3 shows the filter characteristics for the six different STEs of Fig. 3 in the main manuscript. For a, b, and f (please note the different scaling) the characteristics are identical. Only when the conductivity is changed (or the thickness achieving the same result), we observe a shift of the cut-off frequency. For higher conductivity/thickness (e) the cutoff frequency is shifted up, for lower conductivity/thickness (c and d) it is shifted down.

\begin{figure*}
    \includegraphics[width=18cm]{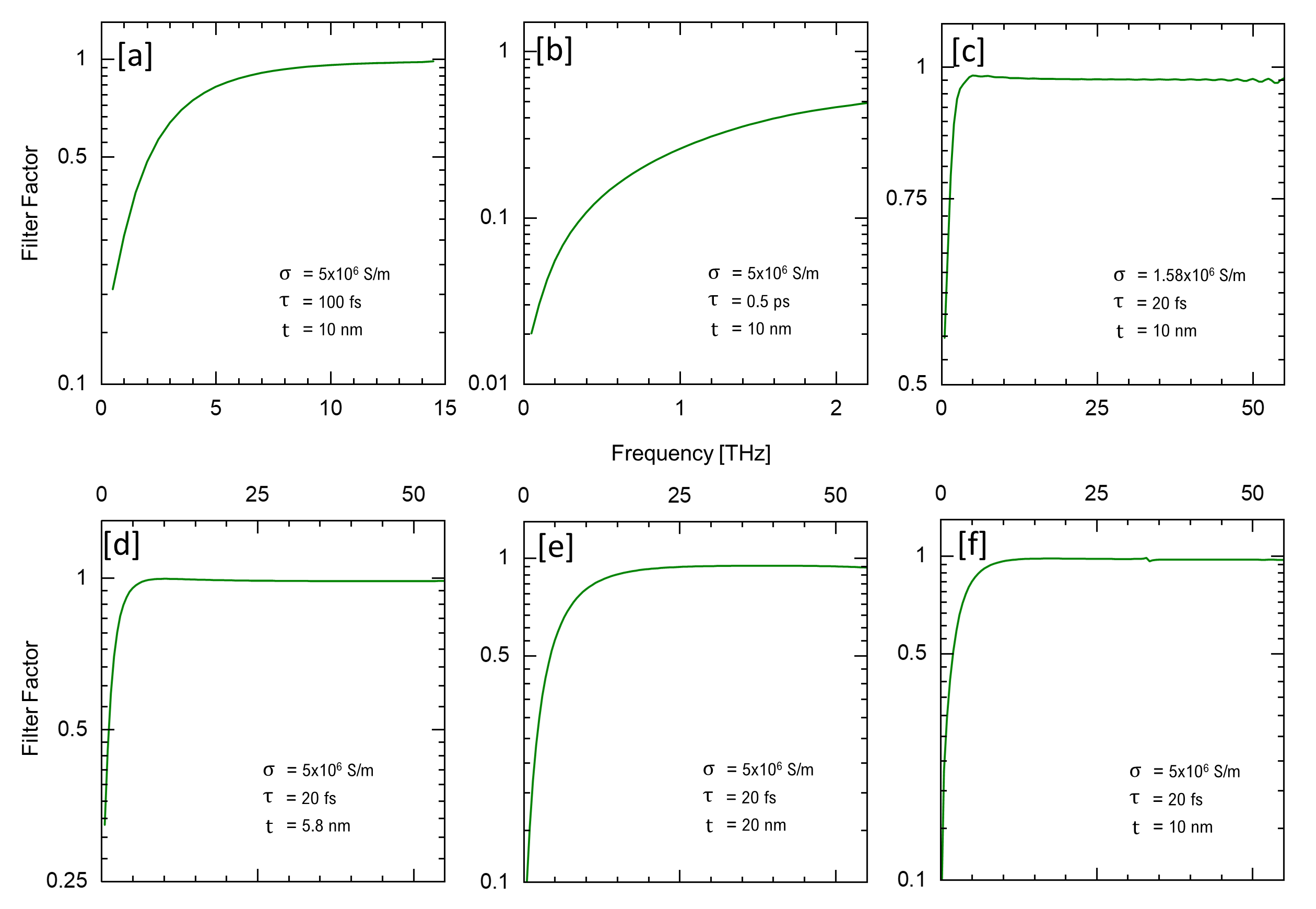}
    \caption{Filter characteristics for the six different STE experiments from Fig.3 in the main manuscript. (a) and (b) show the results for a 100 fs wide and a 0.5 ps wide pulse $I_0$, respectively, using $STE_{ref}$. (c) Shows the result using $STE_{ref}$ with a conductivity reduced to $1.58\times 10^6$ and a 20 fs pulse. The  cutoff frequency is reduced. (d) shows results for $STE_{ref}$ with a reduced thickness (5.8\,nm) with a similarly reduced cutoff frequency. Decreasing the resistance by making $STE_{ref}$ thicker (e) shifts the cutoff frequency to higher values. (f) shows the result for an antisymmetric pulse using $STE_{ref}$. Because the STE is the same as in (a) and (b), the cutoff frequency is also identical.}
\end{figure*}